\documentclass[aps,prd,preprint,superscriptaddress,tightenlines,nofootinbib]{revtex4}

\usepackage{amsfonts,amsmath,amsthm,amssymb}
\usepackage{mathrsfs}
\usepackage{bm}
\usepackage{color}
\usepackage{epsfig}

\newcommand{\PRE}[1]{{#1}}   

\newcommand{\comment}[1]{}

\newcommand{\el}[1]{\label{#1}}
\newcommand{\er}[1]{\eqref{#1}}

\newcommand{\ci}[1]{}

\newcommand{\ke}{\rangle}
\newcommand{\br}{\langle}

\newcommand{\p}{\partial}

\newcommand{\ba}{\begin{eqnarray}}
\newcommand{\ea}{\end{eqnarray}}
\newcommand{\be}{\begin{equation}}
\newcommand{\ee}{\end{equation}}
\newcommand{\bay}[1]{\left(\begin{array}{#1}}
\newcommand{\eay}{\end{array}\right)}

\def\xs{{\sigma}}

\def\xt{{\theta}}

\def\CD{{\cal D}}

\def\CR{{\cal R}}

\begin{document}

\preprint{
\hfil
\begin{minipage}[t]{3in}
\begin{flushright}
\vspace*{.4in}
MPP--2011--41\\
LMU-ASC 13/11\\
CERN-PH-TH/2011-079\\
\end{flushright}
\end{minipage}
}

\vspace{1cm}

\title{Stringy origin of Tevatron $\bm{W jj}$ anomaly
\PRE{\vspace*{0.3in}} }

\author{Luis A. Anchordoqui}
\affiliation{Department of Physics,\\
University of Wisconsin-Milwaukee,
 Milwaukee, WI 53201, USA
\PRE{\vspace*{.1in}}
}

\author{Haim Goldberg}
\affiliation{Department of Physics,\\
Northeastern University, Boston, MA 02115, USA
\PRE{\vspace*{.1in}}
}

\author{Xing \nolinebreak Huang}
\affiliation{Department of Physics,\\
University of Wisconsin-Milwaukee,
 Milwaukee, WI 53201, USA
\PRE{\vspace*{.1in}}
}

\author{Dieter L\"ust}
\affiliation{Max--Planck--Institut f\"ur Physik\\
 Werner--Heisenberg--Institut,
80805 M\"unchen, Germany
\PRE{\vspace*{.1in}}
}

\affiliation{Arnold Sommerfeld Center for Theoretical Physics\\
Ludwig-Maximilians-Universit\"at M\"unchen,
80333 M\"unchen, Germany
\PRE{\vspace{.1in}}
}

\affiliation{Physics Division (TH), CERN                                                    
1211 Geneva 23, Switzerland}

\author{Tomasz R. Taylor}
\affiliation{Department of Physics,\\
  Northeastern University, Boston, MA 02115, USA \PRE{\vspace*{.1in}}
}

\date{April 2011} \PRE{\vspace*{.5in}} \begin{abstract}\vskip 3mm \noindent The invariant mass distribution of dijets produced in association with $W$ bosons, recently observed by the CDF Collaboration at Tevatron, reveals an excess in the dijet mass range $120-160~{\rm GeV/c}^2$, $3\sigma$ beyond Standard Model expectations. We show that such an excess is a generic feature of low mass string theory, due to the production and decay of a leptophobic $Z'$, a singlet partner of $SU(3)$ gluons coupled primarily to the $U(1)$ baryon number. In this framework, $U(1)$ and $SU(3)$ appear as subgroups of $U(3)$ associated with open strings ending on a stack of 3 D-branes. In addition, a minimal model contains two other stacks to accommodate the electro-weak $SU(2)\subset U(2)$ and the hypercharge $U(1)$. Of the three $U(1)$ gauge bosons, the two heavy $Z'$ and $Z''$ receive masses through the Green-Schwarz mechanism. For a given $Z'$ mass, the model is quite constrained. Fine tuning three of its
free parameters is just sufficient to simultaneously ensure: a small $Z-Z'$ mixing in accord with the stringent LEP data on the $Z$ mass; very small (less than 1\%) branching ratio into leptons; and a large hierarchy between $Z''$ and $Z'$ masses.  The heavier neutral gauge boson $Z''$ is within the reach of LHC.
\end{abstract}

\maketitle

It appears that in the last year of the Tevatron's endeavors, it has
pierced the Standard Model's resistant
armor~\cite{:2007qb,Aaltonen:2011mk}.  The latest foray is an excess
at $M_{jj} \simeq 140~{\rm GeV}$ in the dijet system invariant mass
distribution of the associated production of a $W$ boson with 2 jets
(hereafter $Wjj$ production)~\cite{Aaltonen:2011mk}.  The CDF
Collaboration fitted the excess to a Gaussian and estimated its
production rate to be $\sim 4$~pb. This is roughly 300 times the
Standard Model Higgs rate $\sigma(p\bar p \to WH) \times {\rm BR}(H
\to \bar bb)$. For a search window of $120 - 200~{\rm GeV}$, the
excess significance above Standard Model background (including
systematic uncertainties) is $3.2 \sigma$~\cite{Aaltonen:2011mk}.

The CDF $Wjj$ anomaly has been related to the technipion of a low mass
technicolor~\cite{Eichten:2011sh}, to resonant super-partner production in a supersymmetric model with $R$-parity violation~\cite{Kilic:2011sr}, and to a leptophobic $Z'$ gauge
boson~\cite{Bai:2010dj,Buckley:2011vc,Yu:2011cw,Cheung:2011zt}. The suppressed coupling to
leptons in the latter is required to evade the strong constraints of
the Tevatron $Z'$ searches in the dilepton
mode~\cite{Acosta:2005ij}. All existing dijet-mass searches at the
Tevatron are limited to $M_{jj} > 200~{\rm GeV}$~\cite{Abe:1995jz} and
therefore cannot constrain the existence of a $Z'$ with $M_{Z'} \simeq
140~{\rm GeV}$. The strongest constraint on a light leptophobic $Z'$
comes from the dijet search by the UA2 Collaboration, which has placed
a 90\% CL upper bound on $\sigma \times {\rm BR}(Z' \to jj)$ in this
energy range~\cite{Alitti:1990kw}. In this Letter we show that a $Z'$
that can explain the $Wjj$ excess and is in full agreement with
exisitng limits on $Z'$ coupling to quarks and leptons can materialize
in the context of D-brane TeV-scale string compactifications.

At the time of its formulation and for years thereafter, superstring
theory was regarded as a unifying framework for Planck-scale quantum
gravity and TeV-scale Standard Model physics. Important advances
were fueled by the realization of the vital role played by
D-branes  in connecting string theory to
phenomenology.  This has permitted the
formulation of string theories with compositeness setting in at TeV
scales and large extra dimensions~\cite{Antoniadis:1998ig}. 

TeV-scale superstring theory provides a brane-world description of the
Standard Model, which is localized on membranes extending in $p+3$ spatial
dimensions, the so-called D-branes~\cite{Blumenhagen:2001te,Honecker:2004kb}. Gauge
interactions emerge as excitations of open strings with endpoints
attached on the D-branes, whereas gravitational interactions are
described by closed strings that can propagate in all nine spatial
dimensions of string theory (these comprise flat parallel
dimensions extended along the $(p+3)$-branes and transverse
dimensions)~\cite{Blumenhagen:2006ci}. The apparent weakness of gravity at energies below few
TeV can then be understood as a consequence of the gravitational
force ``leaking'' into the transverse compact dimensions of spacetime.

There are two peerless phenomenological consequences for TeV-scale
D-brane string physics: the emergence of Regge recurrences at parton
collision energies $\sqrt{\hat s} \sim M_s \equiv$~string scale, most
distinctly manifest in the $\gamma$ + jet~\cite{Anchordoqui:2007da}
and dijet~\cite{Anchordoqui:2008di} spectra resulting from their decay\footnote{The recent search for string resonances in $pp$-collisions 
by the CMS collaboration~\cite{Khachatryan:2010jd}
at the LHC now excludes a string scale below 2.5 TeV. From the string theory point of view,
D-brane models with a TeV string scale can be obtained by the compactification on special
Calabi-Yay spaces (Swiss cheese manifolds).}; and
the presence of one or more additional $U(1)$ gauge symmetries, beyond
the $U(1)_Y$ of the Standard Model. The latter follows from the
property that the gauge group for open strings terminating on a stack
of $N$ identical D-branes is $U(N)$ rather than $SU(N)$ for
$N>2$. (For $N=2$ the gauge group can be $Sp(1)$ rather than $U(2)$.)

To develop our program in the simplest way, we will work within the
construct of a minimal model in which we consider scattering processes
which take place on the (color) $U(3)$ stack of D-branes. In the
bosonic sector, the open strings terminating on this stack contain, in
addition to the $SU(3)$ octet of gluons $g_\mu^a$, an extra $U(1)$
boson ($C_\mu$, in the notation of~\cite{Berenstein:2006pk} -  this model was also discussed in \cite{Anastasopoulos:2006da}), most
simply the manifestation of a gauged baryon number symmetry. The
$U(1)_Y$ boson $Y_\mu$, which gauges the usual electroweak hypercharge
symmetry, is a linear combination of $C_\mu$, the $U(1)$ boson $B_\mu$
terminating on a separate $U(1)$ brane, and perhaps a third additional
$U(1)$ field $X_\mu$ sharing a $U(2)$ stack  which is also a terminus for the
$SU(2)_L$ electroweak gauge bosons
$W_\mu^a$~\cite{Antoniadis:2000ena}. Any vector boson $Z_\mu'$,
orthogonal to the hypercharge, must grow a mass  in order to
avoid long range forces between baryons other than gravity and Coulomb
forces. The anomalous mass growth allows the survival of global baryon
number conservation, preventing fast proton decay~\cite{Ghilencea:2002da}.

In the minimal $U(3) \times Sp(1) \times U(1)$ D-brane model, the hypercharge \begin{equation} Q_Y \equiv \frac{1}{6} Q_{U(3)} - \frac{1}{2} Q_{U(1)}  \end{equation} is anomaly free. However, the $Q_{U(3)}$ (gauged baryon number) is not anomaly free and we expect this anomaly to be canceled via a Green-Schwarz mechanism. There is an explicit mass term in the Lagrangian for the new gauge field $-\frac{1}{2} M'^2 Y'_\mu Y'^{\mu}$ whose scale comes from the compactification scheme. The scalar that gets eaten up to give the longitudinal polarization of the $Y'$ is a closed string field and there is no extra Higgs particle~\cite{Ghilencea:2002da}. Following~\cite{Berenstein:2006pk} we take $M'$ as a free parameter of the model and use precision electroweak data to determine its value.  As usual, the  $U(1)$ gauge interactions arise through the covariant derivative \be\el{covderi} \CD_\mu = \p_\mu -i g_1 \, B_\mu \, Q_{U(1)} - i \frac{g_3}{\sqrt{6}} \, C_\mu \, Q_{U(3)} \, ,\ee where $g_1$, $g_2$, and $g_3$ are the gauge coupling constants.
Introducing  $S_P \equiv \sin \xt_P$ and $C_P \equiv \cos \xt_P$, the $U(1)$ fields can be projected into massless and massive directions 
\begin{equation} C_{\mu} = C_P Y_\mu{}' + S_P Y_\mu,\quad B_{\mu} =
  S_P Y_\mu{}' - C_P Y_\mu, \label{CBfields} \end{equation} with \be
\tan\xt_P = \sqrt{\frac 2 3}\frac {g_1} {g_3},\quad {\rm and} \quad
\frac{1}{g_Y^2} = \frac{1}{6 g_3^2} + \frac{1}{4 g_1^2} . \label{XHH}\ee
Substituting (\ref{CBfields}) into (\ref{covderi}) we
obtain \begin{equation} g_{Y'} Q_{Y'} = \frac {g_3} {\sqrt 6}
  C_PQ_{U(3)} + g_1 S_PQ_{U(1)} \, . 
\label{sura}
\end{equation} We note that a
value for $g_{Y'}$ will emerge once a normalization for $Q_{Y'}$ is
adopted. (The second relation in Eq.~(\ref{XHH}) depends on the choice of
normalization for the hypercharge). 
For a Higgs ($Q_{U(3)} = 0$, $Q_{U(1)} = -1$, $Q_Y = -1/2$) with vacuum expectation value \begin{equation} \br H \ke = \bay{c} v \\ 0 \eay, \end{equation} the kinetic term $(D_\mu H)^* (D_\mu H)$ gives gives a mass term \begin{equation} (v, 0) \bay{cc} 
- \frac 1 2 \sqrt{g_2^2 + g_Y^2} Z -g_1 S_P Y' & 0 \\ 0 & 
\frac {g_2^2 - g_Y^2}{2\sqrt{g_2^2 + g_Y^2}} Z-g_1 S_P Y' 
\eay ^2 \bay{c} v 
\\ 0 \eay = (\overline M_Z Z +g_1 S_P v Y')^2 , \label{massterM}\end{equation} where \be\el{covderiY}\CD_\mu = \p_\mu -i \frac 1 {\sqrt{g_2^2 + g_Y^2}} Z_\mu (g_2^2 T^3 - g_Y^2 Y) -i g_{Y'} Y_\mu{}' Q_{Y'}  \,, \ee with $T^3 = \xs^3/ 2$ and $g_{Y'} Q_{Y'}$ given in Eq.~(\ref{sura}).  Equation~(\ref{massterM}) together with the mass term $\frac{1}{2} M'^2 Y'{}^2$ lead to a mass matrix \begin{equation}
  \frac{1}{2} (Z, Y') \bay{cc} \overline M_Z^2 & \overline M_Z g_1 S_P v \\
  \overline M_Z g_1 S_P v & g_1^2S_P^2 v^2 +M'^2 \eay  \bay{c} Z \\ Y' \eay = \frac{1}{2} (\overline M_Z Z + g_1 v S_P Y')^2 + \frac{1}{2} M'^2 Y'^2 \, ,  \end{equation}
where $2 \overline M_Z^2 = g_2^2 v^2+ g_Y^2 v^2$ is the usual tree level formula for the mass of the $Z$ particle in the electroweak theory, before mixing~\cite{Berenstein:2006pk}. When the theory undergoes electroweak symmetry breaking, because $Y'$ couples to the Higgs, one gets additional mixing. 
However, to avoid conflict with precision measurements at LEP throughout this Letter we will enforce  negligible $Z-Z'$ mixing and consider $M' \simeq M_{Z'}$~\cite{Umeda:1998nq}.  A comprehensive study of the $M'$ parameter space has been carried out in~\cite{Berenstein:2008xg}, concluding that gauge bosons with $M_{Z'}< 700~{\rm GeV}$ are excluded by the $Z$-pole data from LEP.

In the $U(3) \times U(2) \times U(1)$ D-brane model the $Q_{U(1)}$, $Q_{U(2)}$,
$Q_{U(3)}$ content of the hypercharge operator, \be\el{hyperchargeY} Q_Y =
c_3 Q_{U(3)} + c_2 Q_{U(2)} + c_1 Q_{U(1)}  \,,\ee is not uniquely
determined by the anomaly cancellation requirement. In fact as seen
in~\cite{Antoniadis:2000ena}, there are 5 possibilities. This final
choice does not depend on further symmetry considerations;
in~\cite{Antoniadis:2000ena} it was fixed by requiring partial
unification ($g_3 = g_2$) and acceptable value of $\sin^2 \theta_W$ at
$M_s \sim 6 - 8~{\rm TeV}.$ We take
$c_1 = 0,$ $c_2 = 1$, and $c_3 =-2/3$~\cite{Lust:2008qc}.   For this hypercharge embedding, conventional logarithmic running of coupling constants predicts a high string scale, $M_s \sim {\cal O} (10^{10}~{\rm TeV})$~\cite{Anastasopoulos:2006da}. However, it is possible that threshold corrections in the form of power law running can lower the string scale to the $5 - 10$~TeV region~\cite{Dienes:1998vh} . In what follows we assume this to be the case. The chiral fermion spectrum\footnote{Note that we are considering D-brane quivers, where the right-handed up-quarks $U_i$ originate from open strings ending on the $U(3)$ stack and its orientifold image. This leads to
an antisymmetric matter representation with $Q_{U(3)}=2$. Furthermore, the nomalization of $Q_Y$ for all matter fields in Table~\ref{t:spectrum} is different from that in Eq.(1) by a factor of two.}
is summarized in Table~\ref{t:spectrum}.

\begin{table}
\caption{Chiral fermion spectrum of the $U(3) \times U(2) \times U(1)$ D-brane model.}
\begin{tabular}{c|ccccccc}
\hline
\hline
 Name &~~Representation~~& ~$Q_{U(3)}$~& ~$Q_{U(2)}$~ & ~$Q_{U(1)}$~ & ~~$Q_{Y}$ ~ &~~ $g_{Y'}Q_{Y'}$~ &~~$g_{Y''} Q_{Y''}$\\
\hline
~~$U_i$~~ & $({\bar 3},1)$ &    $\phantom{-} 2$ & $\phantom{-}0$ & $\phantom{-} 0$ & $-\frac{4}{3}$ & $\phantom{-} 0.265$ & $\phantom{-} 0.867$ \\[1mm]
~~$D_i$~~ &  $({\bar 3},1)$&    $-1$ & $\phantom{-}0$ & $\phantom{-} 1$ & $\phantom{-}\frac{2}{3}$ & $-0.098$ & $-0.444$ \\[1mm]
~~$L_i$~~ & $(1,2)$&    $\phantom{-}0$ &  $-1$ & $\phantom{-}1$ & $-1$ & $-0.004$ &  $-0.138$\\[1mm]
~~$E_i$~~ &  $(1,1)$&  $\phantom{-}0$ & $\phantom{-} 2$ &  $\phantom{-} 0$ & $\phantom{-} 2$ & $\phantom{-} 0.078$ & $\phantom{-} 0.255$\\[1mm]
~~$Q_i$~~ & $(3,2)$& $\phantom{-}1$ & $\phantom{-}1 $ & $\phantom{-} 0$ & $\phantom{-}\frac{1}{3}$ & $\phantom{-} 0.172$ & $\phantom{-} 0.561$ \\[1mm]
\hline
\hline
\end{tabular}
\label{t:spectrum}
\end{table}

The covariant derivative is given by~\cite{Anchordoqui:2010zs}
\be\el{covderi2} \CD_\mu = \p_\mu - i \frac{g_3}{\sqrt{6}} \, C_\mu  \,  Q_{U(3)}   -i \frac{g_2}{2} \, X_\mu \, Q_{U(2)}  -i g_1 \, B_\mu \, Q_{U(1)} \, .\ee
The fields $C_\mu, X_\mu, B_\mu$ are related
to $Y_\mu, Y_\mu{}'$ and $Y_\mu{}''$ by a rotation matrix, 
\begin{equation} 
\CR=
\left(
\begin{array}{ccc}
 C_\theta C_\psi  & -C_\phi S_\psi + S_\phi S_\theta C_\psi  & S_\phi
S_\psi +  C_\phi S_\theta C_\psi  \\
 C_\theta S_\psi  & C_\phi C_\psi +  S_\phi S_\theta S_\psi  & - S_\phi
C_\psi + C_\phi S_\theta S_\psi  \\
 - S_\theta  & S_\phi C_\theta  & C_\phi C_\theta 
\end{array}
\right) \,, 
\end{equation}
with Euler angles $\theta$, $\psi,$ and $\phi$. Equation~(\ref{covderi2}) can be rewritten in terms of $Y_\mu$, $Y'_\mu$, and
$Y''_\mu$ as follows
\begin{eqnarray}
\CD_\mu & = & \partial_\mu -i Y_\mu \left(-S_\xt g_1 Q_{U(1)} + \frac 1 2 C_\theta S_\psi g_2 Q_{U(2)} + \frac 1 {\sqrt{6}} C_\theta C_\psi g_3 Q_{U(3)} \right) \nonumber \\  
 & - & i Y'_\mu \left[ C_\theta S_\phi  g_1 Q_{U(1)} + \frac{1}{2} \left( C_\phi C_\psi + S_\theta S_\phi S_\psi \right) g_2 Q_{U(2)} + \frac{1}{\sqrt{6}} (C_\psi S_\theta S_\phi - C_\phi S_\psi) g_3 Q_{U(3)} \right] \\
& - & i Y''_\mu \left[ C_\theta C_\phi g_1 Q_{U(1)} + \frac 1 2 \left(-C_\psi S_\phi + C_\phi S_\theta S_\psi \right) g_2 Q_{U(2)} + \frac 1 {\sqrt{6}} \left( C_\phi C_\psi S_\theta + S_\phi S_\psi\right) g_3 Q_{U(3)} \right]   \, .  \nonumber 
\label{linda}
\end{eqnarray}
Now, by demanding that $Y_\mu$ has the 
hypercharge $Q_Y$ given in Eq.~\er{hyperchargeY}  we  fix the first column of the rotation matrix $\CR$
\begin{equation}
\bay{c} C_\mu \\ X_\mu \\ B_\mu
\eay = \left(
\begin{array}{lr}
  Y_\mu \, \sqrt{6}c_3g_Y /g_3& \dots \\
  Y_\mu \, 2c_2g_Y/g_2 & \dots\\
   Y_\mu \, c_1g_Y/g_1 & \dots
\end{array}
\right) \, ,
\end{equation}
and we determine the value of the two associated Euler angles 
\begin{equation}
\theta = {\rm arcsin} [c_1 g_Y/g_1] = 0
\end{equation}
and 
\begin{equation}
\psi = {\rm arcsin}  [2 c_2 g_Y/ (g_2 \, C_\theta)] = 1.99 \,, 
\end{equation}
where we have taken $M_Z = 91.1876$,  $g_2 = 0.6596$,
$g_3 = 1.215$. The third Euler angle $\phi$ and the coupling $g_1$  are determined by requiring sufficient suppression ($\alt 1\%$) to leptons and compatibility with  the 90\%CL upper limit reported by the UA2 Collaboration on $\sigma (p\bar p \to Z') \times {\rm
  BR} (Z' \to jj)$  at $\sqrt{s} = 630~{\rm GeV}$. The decay width of $Z' \to f\bar f$ is given by~\cite{Barger:1996kr}
\begin{equation}
\Gamma(Z' \to f \bar f) = \frac{G_F M_Z^2}{6 \pi \sqrt{2}}  N_c C(M_{Z'}^2) M_{Z'} \sqrt{1 -4x} \left[v_f^2 (1+2x) + a_f^2 (1-4x) \right] \, ,
\end{equation}
where $G_F$ is the Fermi coupling constant, $C(M_{Z'}^2) = 1 + \alpha_s/\pi + 1.409 (\alpha_s/\pi)^2 - 12.77 (\alpha_s/\pi)^3$, $\alpha_s = \alpha_s(M_{Z'})$ is the strong coupling constant at the scale $M_{Z'}$, $x = m_f^2/M_{Z'}^2$, $v_f$ and $a_f$ are the vector and axial couplings, and $N_c =3$ or 1 if $f$ is a quark or a lepton, respectively. The parton-parton
cross section in the narrow $Z'$ width approximation is given
by
\begin{equation}
\hat \sigma (q \bar q \to Z') =  K \frac{2 \pi}{3} \, \frac{G_F \, M_Z^2}{\sqrt{2}}  \left[v_q^2 (\phi, g_1)+ a_q^2 (\phi, g_1) \right] \, \delta \left(\hat s - m_{Z'}^2 \right) \,,
\end{equation}
where the $K$-factor represents the enhancement from higher order
QCD processes estimated to be $K \simeq 1.3$~\cite{Barger}. After
folding $\hat \sigma$ with the CTEQ6 parton distribution
functions~\cite{Pumplin:2002vw}, taking $M_{Z'} = 140~{\rm GeV}$,  the branching ratio of electrons to quarks is minimized within the $\phi-g_1$ parameter space, subject to saturation of the 90\%CL upper limit~\cite{Alitti:1990kw},
\begin{equation}
\sigma (p\bar p \to Z') \times {\rm BR} (Z' \to jj) \approx 250~{\rm pb} \, .
\end{equation}
This occurs for
for $\phi = 1.87$ and $g_1 = 0.036$, corresponding to a suppression $\Gamma_{Z' \to e^+ e^-}/\Gamma_{Z'\to q \bar q} \sim 0.5\%$. (This also corresponds to $v_u^2 + a_u^2 = 0.355$, and $v_d^2 + a_d^2 = 0.139$.) The UA2 data has a dijet mass resolution $\Delta M_{jj}/M_{jj} \sim 10\%$~\cite{Alitti:1990kw}. Therefore, at 140 GeV the dijet mass resolution is about 15~GeV. This is much larger than the resonance width, which is calculated to be $\Gamma(Z' \to f \bar f)  \simeq 2~{\rm GeV}.$  All the couplings of the $Y'$ boson are now detemined and contained in Eq.~(\ref{linda}). Numerical values are given in Table~\ref{t:spectrum} under the heading of $g_{Y'} Q_{Y'}$.  The corresponding $Wjj$ production rate at the Tevatron ($\sqrt{s} = 1.96~{\rm TeV}$) mediated through $t$ and $u$ channel quark exchange is found to be $\approx 4~{\rm pb}$, in agreement with observation~\cite{Aaltonen:2011mk} and with the recent estimate of~\cite{Cheung:2011zt}.
The rate for the 
associated production channels $ZZ'$, $\gamma Z'$, and  $Z'Z'$ is down by  factors of approximately 3,  5, and 9, respectively~\cite{Cheung:2011zt}.

The second strong constraint on the model derives from the mixing of the $Z$ and the $Y'$ through their coupling to the two Higgs doublets $H$ and $H'$. The criteria we adopt here to define the Higgs charges is to make the Yukawa couplings ($H \bar u q$ and $H' \bar d q$) invariant under all three $U(1)$'s. This leads to $Q_{U(3)} = 3,$ $Q_{U(2)} = 1$, $Q_{U(1)} = 0$, $Q_Y = -1$ and $Q_{U(3)} = 0,$ $Q_{U(2)} = Q_{U(1)} = 1$, $Q_Y = 1$, for $H$ and $H'$ respectively.\footnote{The Higgs fields $H$ with $Q_{U(3)} = 3$ cannot simply be realized by a single open string in the considered D-brane quiver. It has to be thought as the antisymmetric product of three fundamental representations of $U(3)$. Alternatively we could have choosen the Higgs field $H$ as an open string, corresponding to the bifundamental representation with charges $Q_{U(3)} = 0,$ $Q_{U(2)} = Q_{U(1)} = -1$, $Q_Y = -1$. This minimal supersymmetric Standard Model quiver is also consistent with the constraints from string tadpole cancelation. However, the up-quark Yukawa couplings are then forbidden in string perturbation theory and must be generated through D-instanton effects.}  Here, $\br H \ke = (^{v_u}_{0})$, $\br H' \ke = (^{ 0}_{v_d}),$
   $v = \sqrt{v_u^2 + v_d^2} = 172~{\rm GeV}$, and $\tan \beta \equiv
v_u/v_d$~\cite{Antoniadis:2002qm}. To account for $Y''$ we introduced a second term in (\ref{covderiY}), $\CD_\mu = \p_\mu ... -i g_{Y'} Y_\mu{}' Q_{Y'} -i g_{Y''})
Y_\mu{}'' Q_{Y''}$, which is convenient to write as 
\begin{equation}
-i \frac {x_H}{ v_u} \overline M_Z Y_\mu{}' - i \frac{y_H} {v_u} \overline M_Z Y_\mu{}'' + H \to H' \ ,
\end{equation}
where for the two Higgs doublets 
\begin{equation}
x_{H} = -0.252 C_\phi + 1.886 \, g_1 \, S_\phi, \quad  \quad x_{H'} = 2.817 C_\phi
\end{equation}
and 
\begin{equation}
y_{H} = 1.886 \, g_1 \, C_\phi + 0.252 S_\phi, \quad \quad y_{H'} = -2.817 S_\phi \, . 
\end{equation}
The Higgs field  kinetic term together with the Green-Schwarz mass terms  ($-\frac{1}{2} M'^2 Y'_\mu Y'^\mu - \frac{1}{2} M''^2 Y''_\mu Y''^\mu$) yield the following mass square matrix
$$ \bay{ccc} \overline M_Z^2 & \overline M_Z^2 (x_{H} C_\beta^2 + x_{H'} S_\beta^2) & \overline M_Z^2 (y_{H}  C_\beta^2 + y_{H'} S_\beta^2) \\ \overline M_Z^2 (x_{H} C_\beta^2 + x_{H'} S_\beta^2) &
M_Z^2 (C_\beta^2 x_{H}^2 + S_\beta^2 x_{H'}^2) + M'^2 & \overline M_Z^2
(C_\beta^2 x_{H} y_{H} + S_\beta^2  x_{H'} y_{H'})  \\
\overline M_Z^2 (y_{H} C_\beta^2 + y_{H'} S_\beta^2) & \overline M_Z^2 (C_\beta^2 x_{H} y_{H} + S_\beta^2 x_{H'} y_{H'}) & \overline M_Z^2 (y_{H}^2 C_\beta^2 + y_{H'}^2 S_\beta^2) + M''^2\eay \, ,$$ where $x_{H} = 0.139$, $x_{H'} = -0.824,$ $y_{H} = 0.221$, and $y_{H'} = -2.694$.  The free parameters are $\tan \beta$, $M_{Z'},$ and $M_{Z''}$ which will be fixed  by requiring  the shift of the $Z$ mass to lie within 1 standard deviation of the experimental value and $M_{Z'} = 140 \pm 2~{\rm GeV}$. We are also minimizing $M_{Z''}$  to ascertain whether it can be detected at existing colliders.
This leads to $\tan \beta =0.4$, $M_{Z'} \simeq M' \simeq 140~{\rm GeV},$ and $M_{Z''} \simeq M'' \geq 3~{\rm TeV}$.

We now explore (at the parton level) prospects for searches of $Z''$ signals at the Large Hadron Collider (LHC).  All the couplings of the $Y''$ boson are given in Table~\ref{t:spectrum} under the heading of $g_{Y''} Q_{Y''}$. Using these figures we determine $\Gamma_{Z'' \to e^+ e^-}/\Gamma_{Z''\to q \bar q} \sim 0.7\%$. We therefore consider the standard bump-hunting procedure for dijet searches.
We calculate a signal-to-noise ratio, with the signal
rate estimated in the invariant mass window $[M_{Z''} - 2 \Gamma, \, M_{Z''} +
2 \Gamma]$. The noise is defined as the square root of the number of QCD 
background events in the same dijet mass interval for the same
integrated luminosity. As an illustration, we  take $M_{Z''} = 3~{\rm TeV}$, for which 
$\Gamma(Z'' \to f \bar f) =  493~{\rm GeV}$.  For 10~fb$^{-1}$ of data collected at $\sqrt{s} = 14~{\rm TeV}$, we obtain a signal-to-noise ratio of $15\sigma$. 

An obvious question is whether the existing data allow determination of the string mass scale. The anomalous mass contributions to $M_{Z'}$ and $M_{Z''}$ are  proportional (with computable coefficients~\cite{Antoniadis:2002cs}) to $g_{Y'} M_s$ and $g_{Y''} M_s$, respectively. However,  existing data can only determine the products  $g_{Y'} Q_{Y'}$ and $g_{Y''} Q_{Y''}$, see Table~\ref{t:spectrum}. Therefore, a separate measurement of the different quark flavor charges (e.g., by tagging on $b$'s and $t$'s in $Z''$ decays) is necessary to determine the absolute normalization of the couplings and predict the string mass scale.

To summarize, we have considered a low-mass string compactification in which the Standard Model gauge multiplets originate in open strings ending on 3 D-branes. For the non-abelian $SU(3)$ and $SU(2)$ groups the D-brane construct requires the existence of two additional $U(1)$ bosons coupled to baryon number and to the trace of the $SU(2)$ multiplets, respectively.  One linear combination of the three $U(1)$ gauge bosons is identified as the the hypercharge $Y$ field, coupled to the anomaly free hypercharge current. The two remaining linear combinations ($Y', Y''$) of the three $U(1)$'s are coupled to anomalous currents, and grow masses in accord with the Green-Schwarz mechanism. After electroweak breaking, mixing with the third component of isospin results in the three observable gauge bosons, where with small mixing $Z'\simeq Y', \, Z''\simeq Y''$.

For a fixed $M_{Z'}$, the model contain several free parameters -- a single mixing angle and a gauge coupling constant unconstrained by the data -- which are chosen to supress the branching of $Z'$ decay into leptons and to accommodate the UA2 90\%CL data on $p \bar p \to jj X$. The remaining two parameters -- $\tan \beta$ and $M_{Z''}$ -- serve to limit the mass shift (due to mixing) of the electroweak $Z$ to conform with LEP observations. The heavier neutral gauge boson $Z''$ is within the reach of LHC.

In closing, we note that there are some aspects of the model which can lead to observable consequences even in the absence of a light resonant signal. {\em (1)}~The chiral nature of the couplings in Table~\ref{t:spectrum} implies substantial parity violation. Hence, for $M_{Z'} \agt 400~{\rm GeV}$, the parity violating couplings of the $Z'$ to fermions can generate a $t\bar t$ forward-backward asymmetry in $p \bar p$ collisions. {\em (2)} It was noted in~\cite{Buckley:2011vc} that both the $Wjj$ anomaly and the forward-backward asymmetry observed at the Tevatron can be simultaneously explained by a $Z'$ of $M_{Z'} \simeq 140~{\rm GeV}$ with flavor-violating coupling $g_{utZ'} \sim 0.45.$ In principle  these two conditions can be accommodated in  D-brane constructions by introducing two quark families originating  from strings stretching between two stacks of D-branes, and one family  looping  with both ends of a string attached to the color stack~\cite{Blumenhagen:2001te,Lust:2008qc}.  This can give different charges to $u$ and $t$ quarks.

\section*{Acknowledgements}

L.A.A.\ H.G.\ and T.R.T.\ are supported by the U.S.  National Science
Foundation (NSF Grant PHY-0757598 and PHY-0757959).  D.L.\ likes to
thank the theory department of CERN for its hospitality. The work was
partially supported by the Cluster of Excellence "Origin and Structure
of the Universe", in Munich. Any opinions, findings, and conclusions
or recommendations expressed in this material are those of the authors
and do not necessarily reflect the views of the National Science
Foundation.

\end{document}